# Determination of the properties of a superconducting single crystal FeSe using an EPR spectroscopy

S.I. Bondarenko[1], A.A. Prokhorov[2], N.N. Galtsov[1], V.P. Timofeev[1], V.P. Koverya[1], A.V. Krevsun[1]

[1]B. Verkin Institute for Low Temperature Physics and Engineering of the National Academy of Sciences of Ukraine, Kharkiv 61103, Ukraine

[2]Institute of Physics of the Czech Academy of Sciences, Prague, Czech Republic

E-mail: bondarenko@ilt.kharkov.ua

Using an EPR spectrometer, the properties of single-crystal FeSe were studied in a magnetic field up to 6000 Oe at temperatures from 3.5 K to 8 K in the superconducting state and at temperatures above 8K and up to 25K in the normal state. It was shown that by measuring non-resonant spectrometer signals in zero and weak magnetic fields up to 30 Oe, it is possible to determine critical temperature $T_c$, the superconducting transition width, and the value of the first critical magnetic field $H_{c1}$ in FeSe single-crystal. Resonant EPR signals are observed in fields at 1400 Oe and 3400 Oe, which corresponds to the paramagnetism of doubly ionized iron atoms ($Fe^{2+}$) in the FeSe crystal. Non-resonant EPR signals in a magnetic field up to 6000 Oe at temperatures from 3.5K to 8K are a nonlinear function of the field, and correspond to the superconducting mixed state of the FeSe, but at temperatures above 8K (up to 25K) they are field independent.

## 1. Introduction

Over a long period of time, EPR spectroscopy was used primarily to study dielectrics, non-superconducting metals, and metallic compounds that have paramagnetic centers [1]. The usefulness of its use for studying bulk superconductors was questioned by the work of Azbel and Lifshitz [2]. The emergence of "unusual" high-temperature superconductivity (HTSC) in the 1980s and the need to elucidate its nature sparked renewed interest in the use of EPR to study superconductivity. It turned out that using EPR spectroscopy it is possible not only to accurately determine the charge state of ions in the normal state of HTSC compounds, which are part of yttrium and bismuth cuprate compounds [3,4,5], but also to determine in a contactless manner the temperature dependence of the first critical magnetic field, as the field threshold for the occurrence of a mixed state in such superconductors [6].Until very recently, EPR spectroscopy were less used to study another type of HTSCs, namely, superconductors based on iron compounds. In 2021, we published the results of our studies of the effect of hydrogen on iron-containing Fe-Te-Se single crystals [7], some of which were obtained using an EPR spectroscopy [8]. In particular, we confirmed a slight increase in the critical temperature of this compound after the thermal diffusion of hydrogen. The basic structure of such chalcogenides is the superconductor FeSe. To our knowledge, no detailed study of its superconducting and normal properties using EPR spectroscopy has been conducted to date. The aim of this work was to study the properties of a FeSe single crystal using an EPR spectroscopy over a wide range of magnetic fields at temperatures, both below and above the critical value.

## 2. Setting up experiments

The study was conducted on a FeSe sample in the form of a flat single crystal with dimensions of approximately 2.5×2×0.4 mm$^3$. The measurements consisted of recording the EPR signal from the crystal at magnetic field $H$ in the range of 0–6000 Oe for temperatures from 3.5 K to 25 K using a Bruker X-/Q-band E580 FT/CW ELEXSYS spectrometer. A high-Q resonator ER 4122 SHQE Super X High-Q cavity with the $TE_{011}$ mode was used for the measurements. The FeSe single crystal was positioned between quartz rods with diameters of 4 mm. The main characteristics of the spectrometer are: microwave frequency of 9.407 GHz, radiation power of 0.15 mW, magnetic field modulation frequency of 100 kHz, and modulation amplitude of the alternating magnetic field of 20 Oe. The EPR signal is proportional to the absorption intensity of microwave radiation in the sample being studied. In accordance with accepted practice of recording absorption intensity values as a function of magnetic field and temperature, the experimental EPR signal values in this article also do not have a numerical vertical scale. The exception is Fig. 5 with the image of model EPR signals.

Additionally, measurements of the temperature dependence of the magnetization of the single crystal were performed using a quantum superconducting magnetometer MPMS-XL5 from Quantum Design.

### 3. Results of measurements of FeSe characteristics and their discussion

Before the EPR measurements, the superconducting transition temperature $T_c^{onset}$ of the FeSe single-crystal sample was determined using MPMS-XL5. Figure 1 shows the dependence of the magnetic moment $m$ of the studied single crystal on temperature in the superconducting phase transition region. The sample was placed in a magnetic field $H$ = 5 Oe, directed along the main crystallographic axis $c$ of the single crystal.

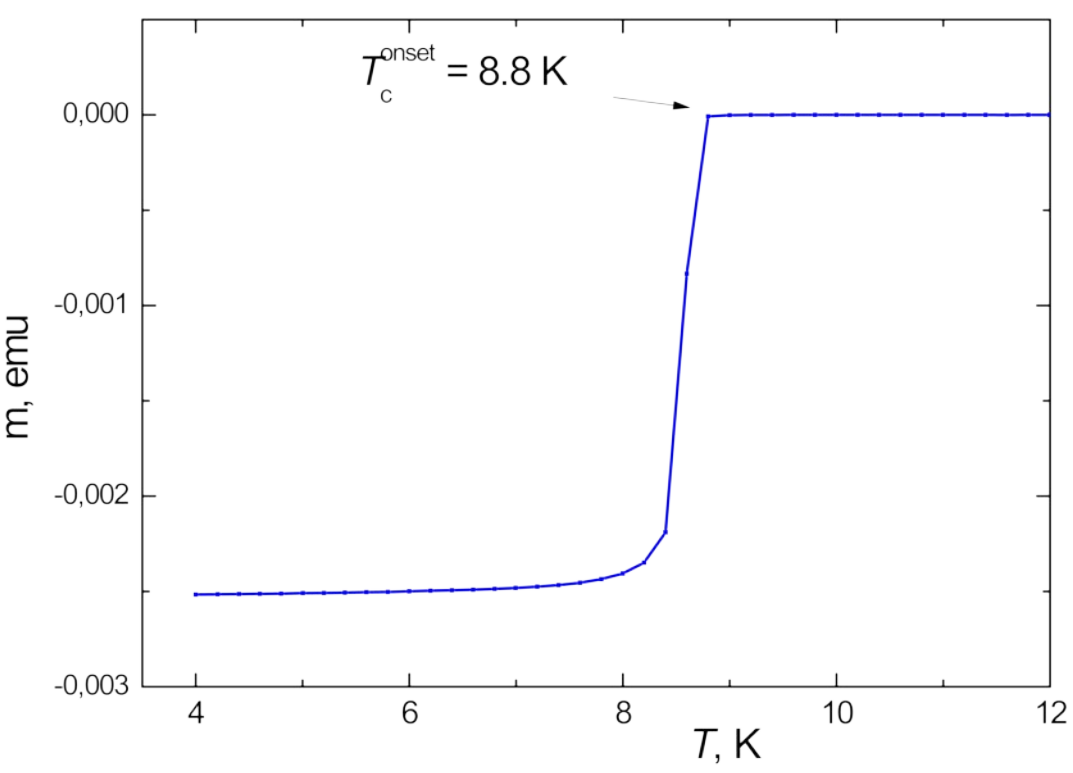

Fig. 1. (Color online) Dependence of the magnetic moment ($m$) of a FeSe single crystal on the temperature.

From the experimental curve $m(T)$, it is evident that the onset temperature of the superconducting transition $T_c^{onset}$ is 8.8 K. Figure 2 shows a family of field dependences of the EPR signal of a FeSe single crystal at various temperatures.

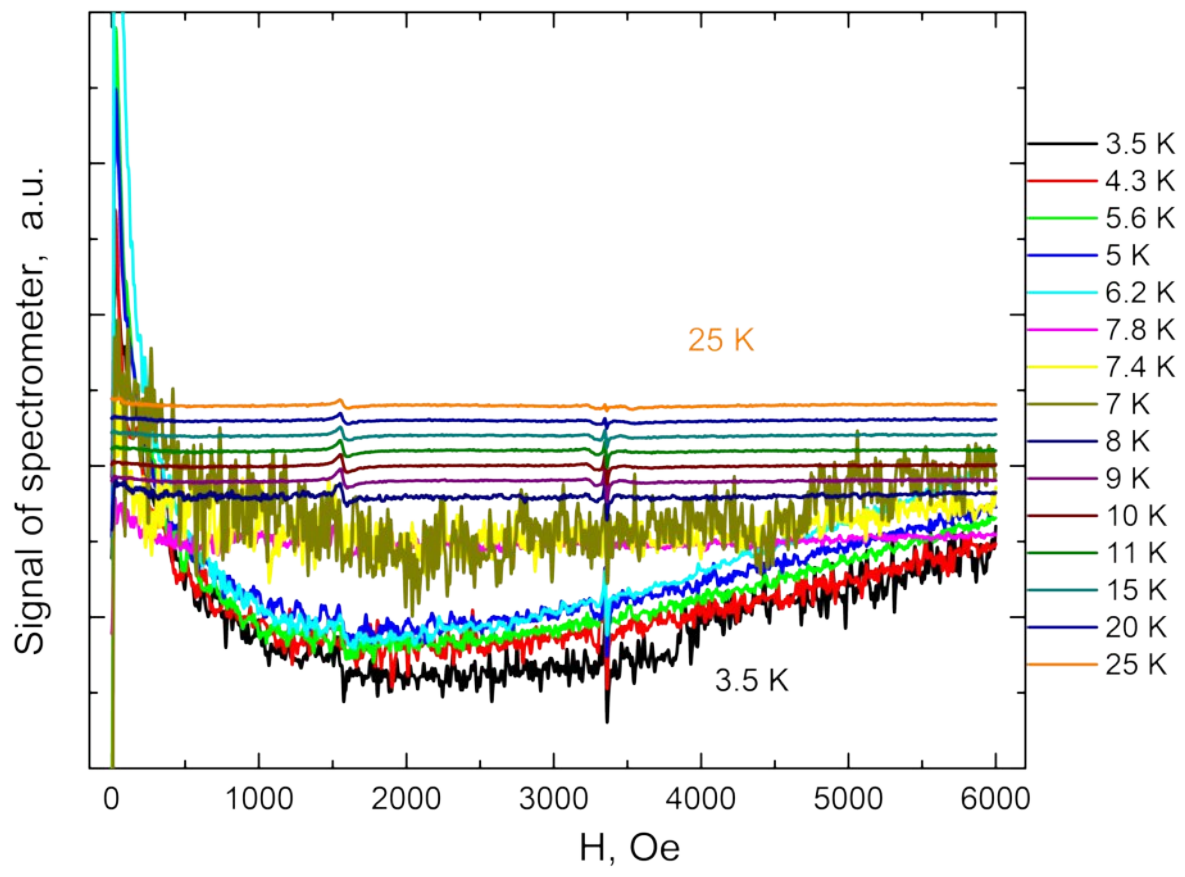

Fig. 2. (Color online) Field dependences of the EPR signal from a FeSe single crystal (in arbitrary units) on the magnitude of the magnetic field $H$ at various temperatures. For ease of analysis, the experimental curves displaying the obtained characteristics are shifted along the vertical axis by equal intervals relative to the dependence at $T$ = 3.5 K.

It can be seen that at temperatures above the critical temperature of the single crystal (about 8 K), the EPR signal is independent of the magnetic field, with the exception of two local features in the form of typical EPR resonances at two field values: 1400 Oe and 3400 Oe. In contrast, at temperatures below the critical temperature, pronounced nonlinear field dependences exist. Near zero magnetic field, the signal increases significantly, and with increasing field, its amplitude decreases to zero and then increases from 3.5 K to the critical value ($T_c^{onset}$). At temperatures below the critical temperature, the EPR resonances in Fig. 2 are poorly distinguishable. The nonlinear field dependences of EPR signal at temperatures from 3.5 K to 25 K have wide minima in the region of magnetic fields of about 2000-2500 Oe and are maintained over the entire range of the used magnetic field (up to 6000 Oe). More detailed behavior of the EPR signal between zero and low magnetic fields up to 30 Oe is shown in Fig. 3.

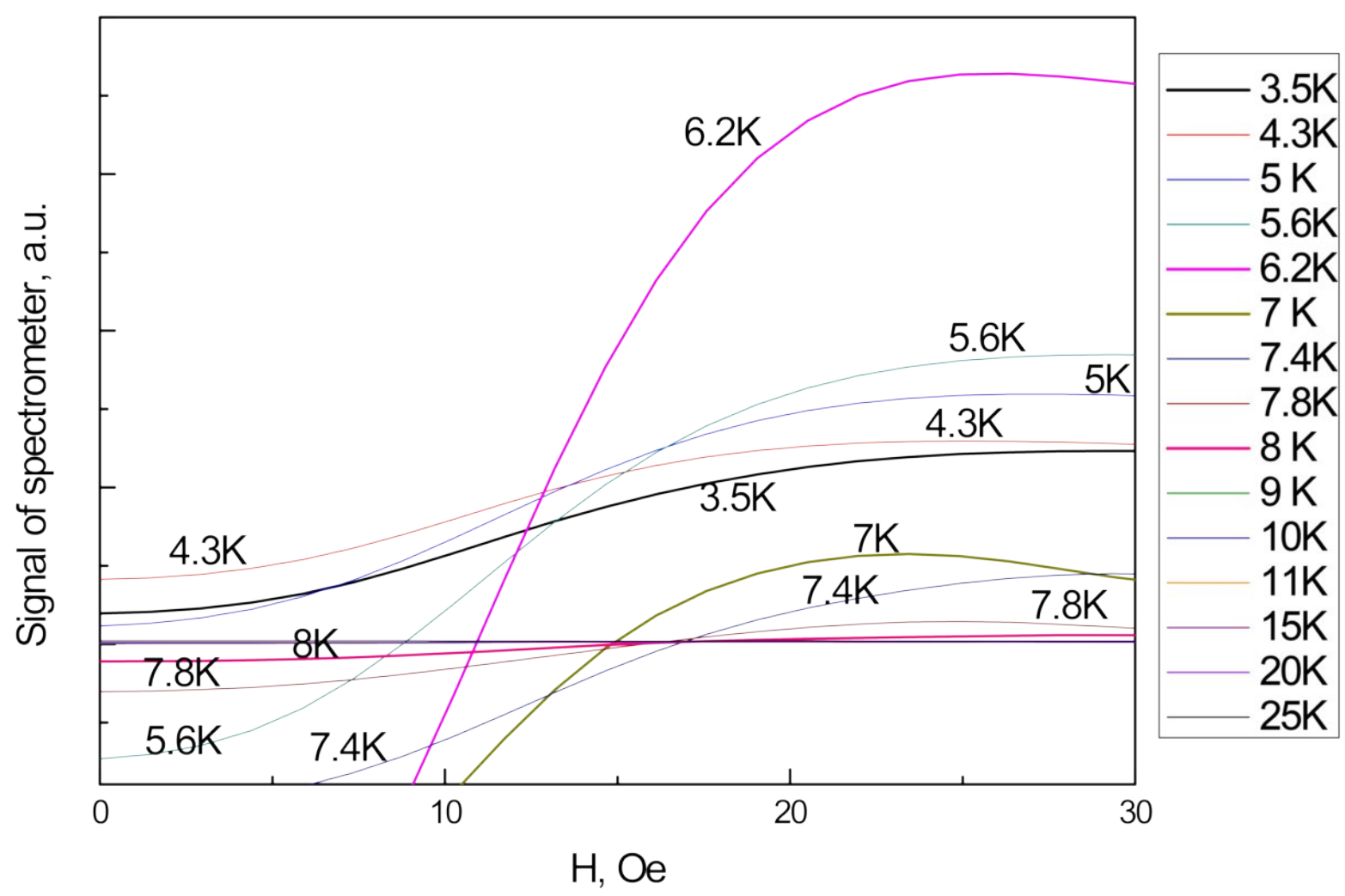

Fig. 3. (Color online) Dependences of the non-resonant EPRsignal of the crystal on the magnetic field, starting from its absence, up to 30 Oe at different temperatures (indicated next to each curve additionally for ease of reading).

Based on the characteristics shown in Fig. 3, the dependence of the EPR signal on temperature $T$ and weak magnetic field $H$ were constructed (Fig. 4).

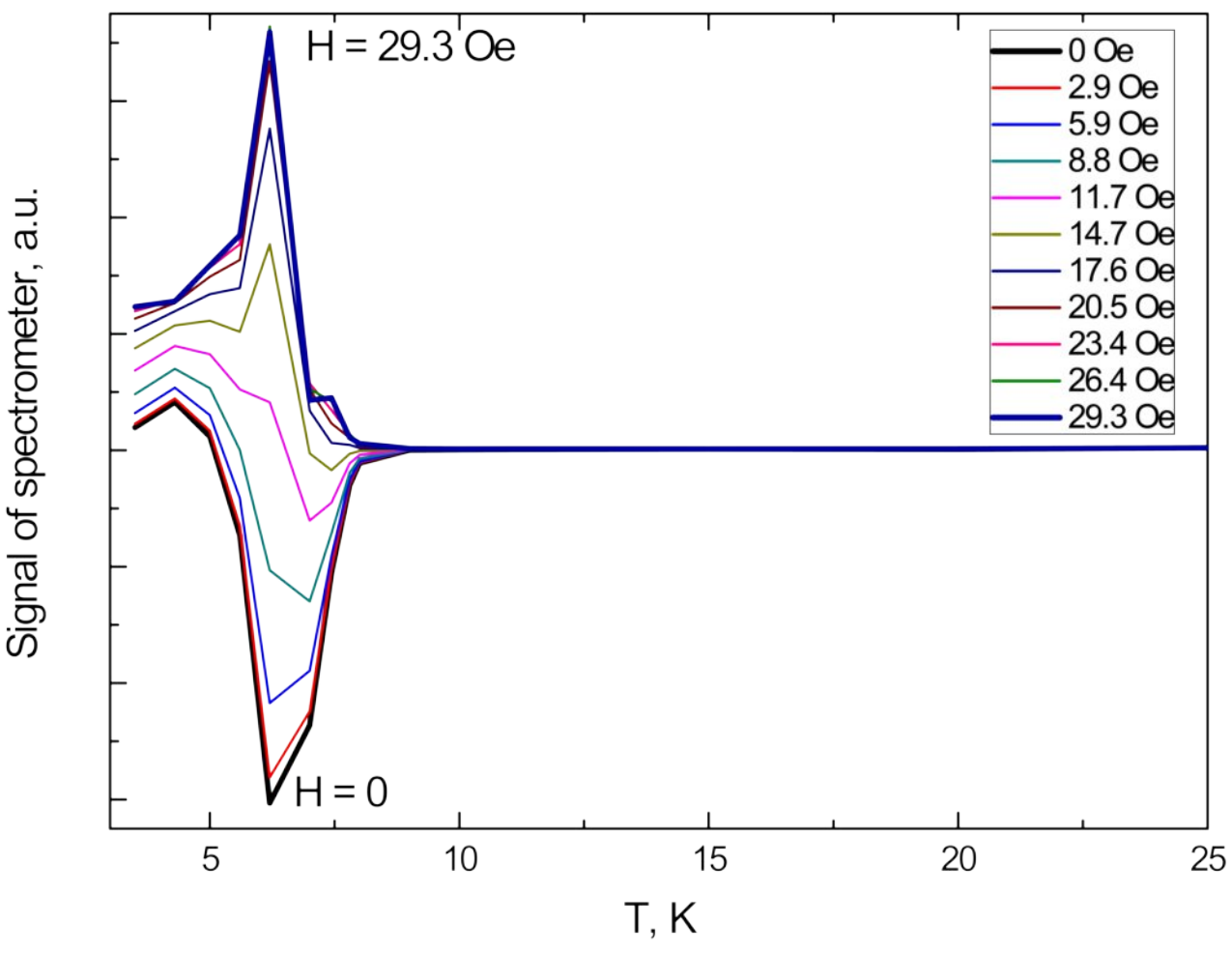

Fig. 4. (Color online) Temperature dependence of the spectrometer output signal when studying a single-crystal FeSe sample in the region of low magnetic fields (0 - 29.3 Oe). The signal is maximum and positive at $H = 29.3$ Oe and is minimum and negative at $H = 0$ Oe.

Several features of the dependences in Figs. 4 can be distinguished. There are negative and positive EPR signals. Negative signals exist in a magnetic field from zero to approximately 12 Oe (more precisely, up to 11.72 Oe), and positive signals in fields from 14 Oe to 29.3 Oe. The transition from negative to positive signals is accompanied by a 180-degree phase change in the EPR signal. The dependences have extrema at a temperature of approximately 6 K. Negative signals vanish at a temperature above to 8 K and at a temperature of below 4.0 K. The largest negative extremum exists in a zero magnetic field. The action of a weak magnetic field on the sample leads to a decrease in negative extrema as the field increases from zero to 11.72 Oe. Conversely, an increase in the magnetic field from 14 Oe to 29.3 Oe leads to an increase in positive extrema.

Let us now discuss the measurement results. When analyzing the EPR signals, it is necessary to take into account the peculiarity of their formation. The resonant EPR signal is proportional to the first field derivative of the microwave absorption power in its resonator [1, 9], and not to the absorption power itself. Consequently, the non-resonant EPR signal is also proportional to the field derivative of the radiation power in the resonator. The non-resonant temperature dependences of the EPR signal shown in Fig. 4 are in the temperature range of 4.0-8 K. As is known, the resistive superconducting trasition of the FeSe compound in zero external magnetic field is located in this temperature range [11]. Consequently, Fig. 4 represents a family of temperature derivatives of this transition, the width of which is approximately 4.0 K. It becomes possible to determine other parameters of the resistive junction based on the features of the EPR signals in Fig. 4. To test this possibility model constructions, we will select two experimental temperature dependences of the EPRsignal in Fig. 4, one at zero magnetic field, and the second for a field of 29.3 Oe.

To compare the experimental dependences with the model representations, we depict two typical temperature dependences of the resistivity ($R^*$) of a certain superconductor (Fig. 5a, curves 1, 2). Such dependences are usually recorded when passing a transport current through a superconductor when the external magnetic field is zero (curve 1) and when an external magnetic field acts on the sample (curve 2).

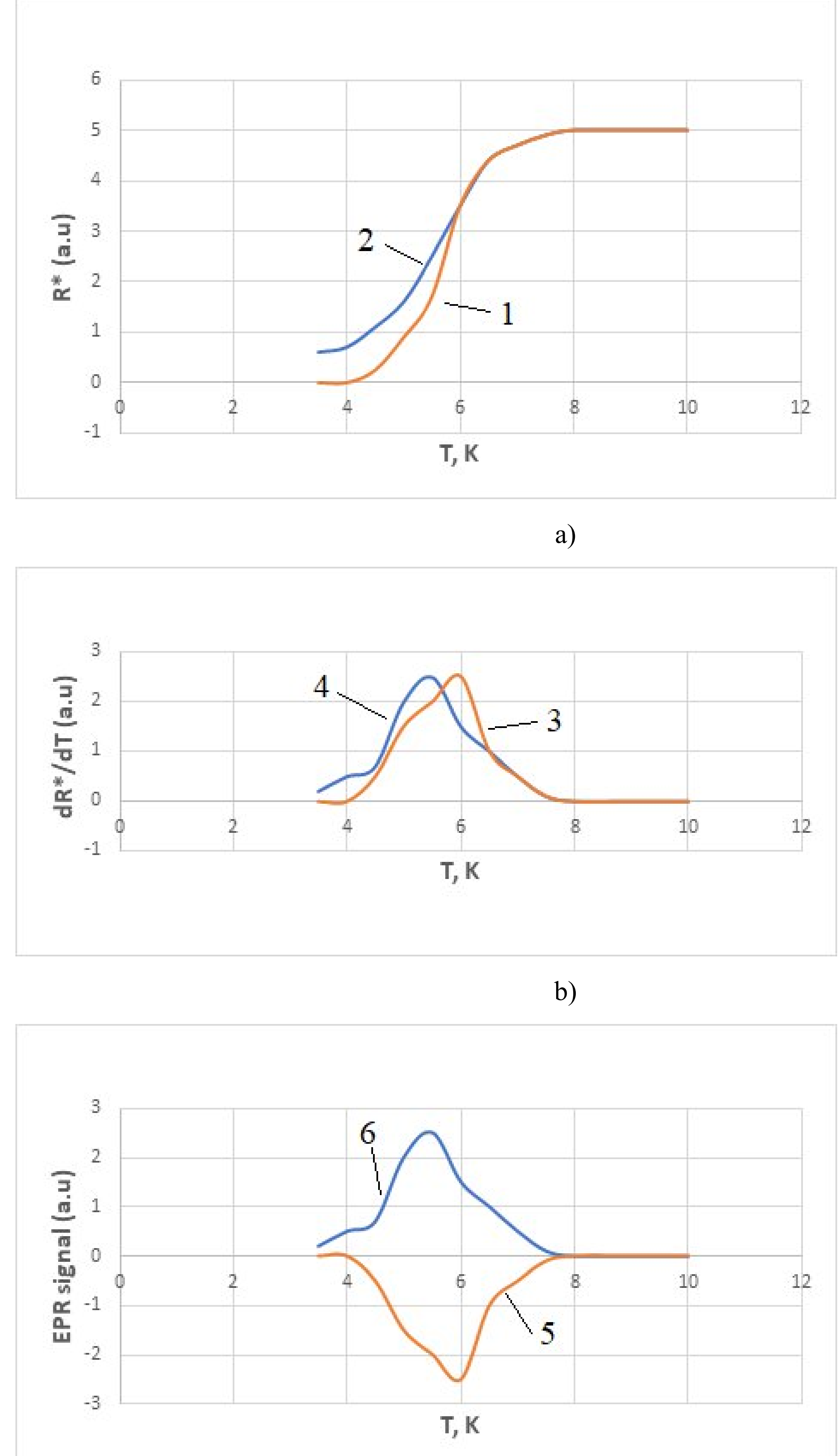

a)

b)

c)

Fig. 5. Temperature dependences of the resistance ($R^*$) of a certain superconductor with a critical temperature of about 8 K (a – curves 1, 2), their first derivatives d$R^*$/d$T$ (b – curves 3, 4), as well as the corresponding calculated temperature dependences of the EPR signal (c – curves 5, 6).

Figure 5b shows the temperature derivatives (curves 3, 4) of the resistivity dependences 1, 2. Derivative 4 is similar in shape and sign to the EPR signal in a field of 29.3 Oe in Figure 4, while derivative 3 is positive, unlike the negative EPR dependence in a zero field in Figure 4. This difference in recording the same superconducting transition process of a sample, in our opinion, is caused by the difference in the principles of measuring the electrical power in the superconducting transition region using a high-frequency EPR spectrometer and by passing a direct current (positive in this case) through the sample while also recording the positive voltage across it using the contact four-probe method (FPM). With FPM, the temperature change in the voltage across the sample is recorded when a constant transport current is passed through it. The electrical power consumed by the sample is converted into heat. The resistance of a sample is determined by dividing the voltage by the current, and the power is equal to the product of the voltage and the current. At temperatures above the critical temperature ($T_c$), the voltage remains virtually unchanged, with only normal electrons serving as current carriers. Upon reaching $T_c$, the voltage and power consumption decrease. According to Gorter's two-fluid model of superconductivity [13], this is caused by the appearance of Cooper pairs of electrons, the density of which in the sample increases as the temperature decreases further, while the voltage and resistance decrease. The current through the sample begins to be carried jointly by normal electrons and Cooper pairs. When a superconducting sample is additionally exposed to an external magnetic field, the superconducting current induced by it is carried solely by Cooper pairs, subtracting them from the transport current. As a result, the proportion of pairs in the transport current decreases, the resistance of the sample increases, and the shape of the superconducting temperature transition changes.

When performing measurements with an EPR spectrometer, a direct transport current is not passed through the superconducting sample, but a high-frequency current is excited in it contactlessly under microwave radiation. To determine the sample's electronic properties in this case, a contactless recording method is required. In an EPR spectrometer, this can be achieved due to the dependence of its signal on the quality factor of the sample-resonator system (SRS). The microwave resonator in a traditional spectrometer is made of non-superconducting materials and is designed to record changes in the SRS quality factor ($Q$) depending on the absorption power of the radiant energy in the non-superconducting sample (with subsequent conversion of this power value into its field derivative). Introducing a superconducting sample with diamagnetic properties into the resonant cavity volume dramatically alters the electromagnetic characteristics of the SRS. Instead of absorption of the microwave radiation in the non-superconducting sample, it is reflected from the superconducting sample. Thus, the electrical power consumption for maintaining resonance in the spectrometer is reduced and, accordingly, the quality factor of the SRS increases instead of decreasing, its resonance and phase characteristics change, as well as the recorded field and temperature derivatives.

If a type-II superconductor is introduced into the resonator, its response to an external magnetic field may become more complex. This is because, upon reaching the first critical magnetic field, such a superconductor ceases to be an ideal diamagnet and, due to the formation of Abrikosov vortices, transforms into a special type of paramagnet. Quantized magnetic flux in the form of Abrikosov vortices with normal cores penetrates the superconductor, leading to the absorption of microwave radiation. This, in turn, can lead to a decrease in the $Q$ factor of the resonator and a change in its other properties, in particular, the signal phase by 180 degrees, which we observed in our experiment (Fig. 4). Thus, unlike a FPM, the recorded signal from a type-II superconductor, such as a FeSe single crystal, can be either negative in the absence of vortices or positive in the presence of vortices, depending on the strength of the external constant magnetic field. In this case, the magnetic field corresponding to a 180-degree phase jump of the spectrometer signal in a field of 12 Oe at a temperature of about 6 K is equal to the first critical field of the single crystal.

Based on the above, it follows that the temperature derivative of the resistance (curve 4 in the FPM in Fig. 5b) in a field exceeding 12 Oe should correspond to the positive values of the EPR signal shown in Fig. 5c (curve 6). The correctness of this assumption is confirmed by the correspondence between the EPR signal in Fig. 5c and the results of the experimental measurements shown in Fig. 4. In contrast, we should (in accordance with the peculiarity of measurements using a spectrometer) arrange the calculated temperature derivative of the resistance(curve 3 in the FPM in Fig. 5b) in zero magnetic field in a mirror image, i.e. with a phase shift of 180 degrees, in the negative part of the scale of the calculated signal of the spectrometer, shown in Fig. 5c (curve 5). Moreover, the model representation of this calculated dependence of the EPR signals completely corresponds to the shape and values of the experimental dependence in a zero magnetic field in Fig. 4.

As can be seen from the comparison of Fig. 5a, 5b, 5c, from the shape of the temperature dependences of the EPR signal within the superconducting transition, it is possible to determine the critical temperature ($T_c^{onset} \approx 8$ K) of the FeSe, the temperature at which the zero value of resistivity is reached when the temperature is reduced to 4.0 K, the width of the superconducting transition (about 4 K), the inflection temperature of the $R^*(T)$ dependence curve corresponding to the negative extremum on the derivative at a temperature of about 6 K, as well as the magnitude of the first critical magnetic field and the shape of the resistive transition of the superconductor in an external magnetic field.

The experimental EPR value of the first critical magnetic field of a FeSe single crystal sample ($H_p$ = 12 Oe) can be compared with its value obtained using the magnetization reversal method [11]. Such a comparison is possible given that this value depends not only on the superconducting properties of the sample but also on its geometric shape, i.e., on the form factor ($N$). Critical fields of samples with $N = 0$ are typically compared. The relationship between $H_p$ and the first critical field $H_{c1}$ of a sample with $N = 0$ is given by the formula [12]:

$$H_p \approx H_{c1} \, (d/w)^{0.5} \, , \qquad (1)$$

where $d$ and $w$ are the thickness (0.4 mm) and width (2 mm), respectively, of the sample. For our sample $H_{c1} \approx 27$ Oe, and for the sample from work [11] $H_{c1} \approx 25$ Oe. The closeness of the $H_{c1}$ values confirms the correctness of the explanation of the 180-degree jump in the phase of the spectrometer signal as a transition of the FeSe crystal to a mixed state.

If necessary, the parameters of the temperature dependences of the superconductor's resistance can be similarly determined using the experimental dependences of the EPR signals at other values of the magnetic field.

Let's analyze the results of EPR resonance measurements of a FeSe single crystal. The observation of resonance features in the field dependences of the EPR signal in magnetic fields of approximately 1400 Oe and 3400 Oe indicates that the crystal has paramagnetic properties, and the resonance field values allow us to determine the specific nature of its electronic state. This specific nature is determined by the $g$-factor of the material being studied, which can be calculated from the relation:

$$\nu = g \mu_0 \mu_B H / h, \qquad (2)$$

where ν, $\mu_0$, $\mu_B$, and $h$ are the EPR frequency, vacuum magnetic susceptibility, Bohr magneton, and Planck constant, respectively. According to [1], for a FeSe single crystal in a magnetic field of 1400 Oe and 3400 Oe, $g \approx 4.8$ and 2.0 were obtained. These values are characteristic of $Fe^{2+}$ ions in various crystallographic environments. They differ significantly from the $g$-factor of pure iron ($g$ = 4). In this case, this means that the iron ions in the FeSe have a charge of +2 ($Fe^{2+}$) and correspond to the $3d^6$ electron state, $S = 2$. Strong orbital interactions and crystal field effects are observed for such ions. For $Fe^{3+}$ ($3d^5$, $S = 5/2$), such high $g$-factor values are not typical, which excludes it as the source of the observed resonance at 1400 Oe [1]. Thus, new information on the nature of the chemical bond between iron and selenium was obtained. The shape of the two detected resonances is clearly visible on the bottom curve of Fig. 6.

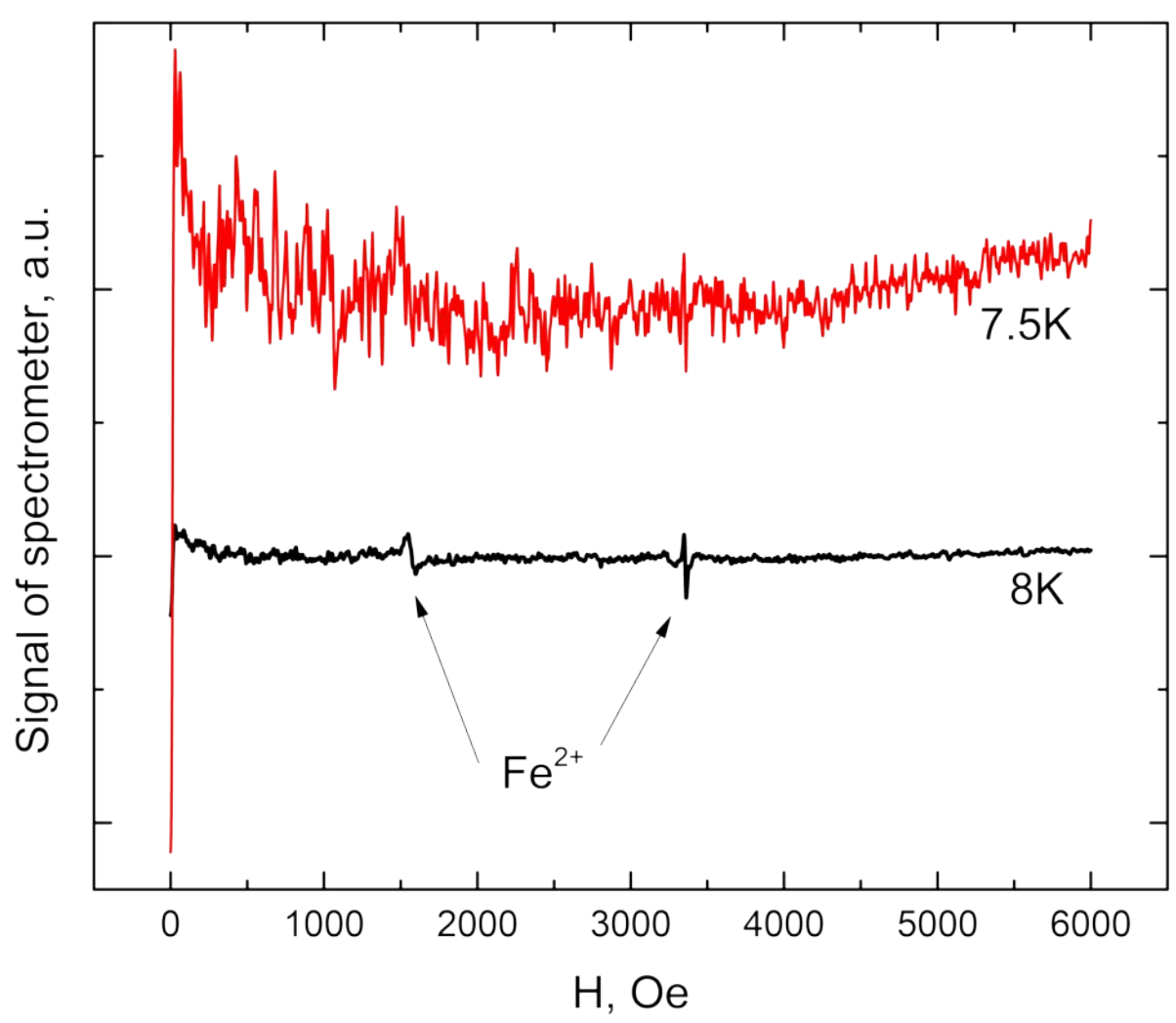

Fig. 6. (Color online) Dependences of the EPR signal intensity on a magnetic field at a temperature of about 8 K for the non-superconducting state of the single crystal (bottom curve) and for its superconducting state at a temperature of 7.5 K (for comparison).

In conclusion, we present an explanation of the nonlinear behavior of the EPR signal dependence on magnetic field (Fig. 2) in the range of 0–6000 Oe at temperatures from 3.5 K to 8K. As can be seen in Figs. 3 and 4, the crystal is in a superconducting state at these temperatures in a magnetic field from zero to 30 Oe. The nonlinearity of the dependences in this magnetic field range is fully explained by the characteristics of the superconducting phase transition of the FeSe crystal. Considering that this crystal is a type-II superconductor with a second critical magnetic field value $H_{c2}$ » 170 kOe, the observed nonlinear magnetic dependence of the EPR signal in fields from 30 Oe to 6 kOe can be explained by the peculiarity of the type of non-resonant EPR signal during magnetization of type-II superconductors by an external magnetic field exceeding $H_{c1}$, and the nonlinear process of penetration of Abrikosov vortices into the superconductor and the formation of a mixed state[14].

## 4. Conclusions

1. Non-resonant EPR signals in FeSe single crystal vary versus magnetic field and temperature in the range $H$ = 0–6000 Oe and $T$ = 3.5–25 K, while EPR resonances exist only at $H$ = 1400 Oe and 3400 Oe.

2. Temperature dependences of non-resonant EPR signals from FeSe single crystal at temperatures of 3.5–8 K in zero and weak magnetic fields allow contactless determination of the temperature and magnetic parameters of its superconducting transition, in particular, the magnitude of the first critical magnetic field ($H_{c1}$) at the transition.

3. The emergence of a mixed superconducting state in a FeSe single crystal in magnetic field is accompanied by a previously unknown sharp change in the phase of the non-resonant EPR signal by 180 degrees.

4. The nonlinear magnetic dependence of the nonresonant EPR signal of the superconducting FeSe single crystal at temperatures from 3.5 K to 8 K in a magnetic field from $H_{c1}$ to 6000 Oe can be explained by the existence of a mixed state of the crystal as a type II superconductor.

5. The resonance EPR signals made it possible to establish the values of the $g$-factors of the FeSe single crystal (4.8 and 2), at which its iron ions have a charge of +2 ($Fe^{+2}$).

The authors express their gratitude to D. A. Chareev for preparing FeSe single crystals, M. I. Kobets and S. N. Shevchenko for useful discussions, and A. V. Dolbin for assistance in organizing the experiments.

This work was financially supported by the of leading Program of the National Academy of Sciences of Ukraine "Fundamental research on the most important problems of natural sciences" (section "Quantum nano-sized superconducting systems: theory, experiment, practical implementation"). State registration number of the work is 0122U001503. V.P.K. was supported by grant of the IEEE Magnetic Society (project No.9918) and National Research Foundation of Ukraine (Grant No.2025.07/0044).

**Referenses**

1. S. A. Altshuler and B. M. Kozyrev, Electron Paramagnetic Resonance, State Publishing House of Physical and Mathematical Literature, M., 1961.

2. M. Ya. Azbel, I.M. Lifshitz, JETP, **33**,792 (1957).

3. D. L. Lyfar', D. P. Moiseev, A. A. Motuz, S. M. Ryabchenko, S. K. Tolpygo, Low Temperature Physics, **13**, No. 8, pp. 503–505 (1987). https://doi.org/10.1063/10.0031770

4. R.I. Khasanov, Yu.M. Vashakidze and Yu.I. Talanov, Physica C, **218**, 51-58(1993). https://doi.org/10.1016/0921-4534(93)90264-Q

5. L. Salakhutdinov, Y. Talanov, E. Giannini, R. Khasanov,Appl.Mag.Res., **70**, № 1, P. 37–46. (2011). https://doi.org/10.1007/s00723-010-0176-2

6. M.K. Aliev, Ya. Vavryshchuk, S.P. Volosyanyj, T.M. Muminov, B.A. Olimov, I. Kholbaev, Fizika Tverdogo Tela, **31** (9), 254-257 (1989).

7. A.I. Prokhvatilov, V.V. Meleshko, S.I. Bondarenko, V.P. Koverya, A. Wisniewski, Low. Temp.Phys., **46**, 181–186 (2020). https://doi.org/10.1063/10.0000538

8. S.I. Bondarenko, A.I. Prokhvatilov, R. Puzniak, J. Pietosa, A.A. Prokhorov, V.V. Meleshko, V.P. Timofeev, V.P. Koverya, D.J. Gawryluk, A. Wisniewski, Materials, **14**, 7900 (2021). https://doi.org/10.3390/ma14247900

9. C.P. Poole, Electron spin resonance, Interscience Publishers, N.Y., 1967.

10. E. Nazarova, N. Balchev, K. Nenkov, K. Buchkov, D. Kolacheva, A. Zachariev and G. Fuchs, Supercond. Sci. Technol., **28**, 025013 (2015). https://doi.org/10.1088/0953-2048/28/2/025013

11. M. Abdel-Hafiez, J. Ge, A. N. Vasiliev, D. A. Chareev, J. Van de Vondel, V. V. Moshchalkov, and A. V. Silhanek, Phys.Rev.B **88**, 174512 (2013). https://doi.org/10.1103/PhysRevB.88.174512

12. E. Zeldov, J. R. Clem, M. McElfresh, and M. Darwin, Phys. Rev. B **49**, 9802 (1994). https://journals.aps.org/prb/abstract/10.1103/PhysRevB.49.9802

13. E. A. Lynton, "Superconductivity," Methuen, London, 1962.

14. D. Saint-James, G. Sarma, E.J. Thomas, Type II superconductivity, Pergamon Press, 1969.